# An Energy Efficient Routing Algorithm for Wireless Sensor Networks Using Mobile Sensors

Meysam Yari [*1], Parham Hadikhani[2], Mohammad Yaghoubi[3], Raza Nowrozy[3], Zohreh Asgharzadeh[4]

[*1] Department of Computer Engineering, Islamic Azad University Shabestar Branch, Shabestar, East Azerbaijan, Iran

meysam.yari68@yahoo.com [1]

[2]Department of Computer Engineering, Pasargad Higher Education Institute, Shiraz, Fars, Iran

[3]College of Engineering and Science, Victoria University, Ballarat Road, Footscray, VIC 3011, Australia

[4]Department of Computer Engineering, Payame Noor University Germi Branch, Germi, Ardabil, Iran

## ABSTRACT

The increasing usage of wireless sensor networks in human life is an indication of the high importance of this technology. Wireless sensor networks have a vast majority of applications in monitoring and care which are known as target tracking. In this application, the moving targets are monitored and tracked in the environment. One of the most important challenges in this area is the limited energy of the sensors. In this paper, we proposed a new algorithm to reduce energy consumption by increasing the load balancing in the network. The proposed algorithm consists of four phases. In the first phase, which is the hole prevention phase, in each cluster, it is checked by the cluster heads that if the energy level in an area of the cluster is less than the threshold, a mobile node is sent to that area. The second phase is the update phase. In this phase, the parameters required to detect a hole are updated. In the third phase, the hole in the cluster is detected, and in the fourth phase, the hole is covered by static or moving nodes. A comparison of simulation results with the well-known and successful routing method in wireless sensor networks show that the proposed method is suitable and working properly.

## 1  Introduction

Wireless sensor networks are known for potential applications such as environmental monitoring and care. Such applications are also known as target tracking, in which moving targets are monitored and tracked. Wireless sensor networks and subsequent target tracking face challenges that overshadow their proper performance. One of the most important challenges are the coverage of sensor nodes on the moving target, the lack of energy due to limited power supply for the sensors and the discharge of the sensors in the area. In these cases, the issues that associated with energy compromise the performance and lifetime of the entire network, the issues that associated with coverage compromise the quality of interception.

Covering is an important debate in the invention of target tracking that affects the quality of monitoring on the operating environment. In general, high coverage means high tracking quality in tracking applications. There are two types of coverage for wireless sensor networks: The first is full coverage over the entire network, which has high power consumption. Because the nodes are always on and therefore the energy is discharged earlier. Second is the spot coverage that just target areas are covered. In this case, the power consumption is reduced because of not covering unnecessary areas. Consequently, the sleep nodes are more than the previous model and the network lifetime is prolonged. In [1-3], they use mobile sensor nodes to cover areas of the network that have been out of reach of nodes for various reasons. The mobile sensor nodes play a key role so that whenever the energy hole is observed, the mobile nodes move to cover the location of the hole. But it is clear that these methods are not applicable to wireless sensor networks with fixed nodes. In 2015, in [4] a method was presented that energy holes were identified and covered in networks with fixed nodes. We then explain this method and combine it with previous methods to obtain the energy balance across the whole network, which increases the lifetime of the network and finally compares the results with [4]. In [4], the authors used sensor nodes that were able to extend their sensory range and had two advantages over their previous methods. First, detection of hole locally (each node individually recognizes the hole and tries to cover the hole). Second, using energy-aware techniques to repair the hole coverage (the node, which has more residual energy and detects the hole, has a higher priority to increase its sensory range.). This article is done in three important phases:

Phase I (update): This phase identifies nodes that are within the sensory range of another node. Therefore, neighbour nodes are aware of each other.

Phase II (coverage hole detection): In this phase, each node sends a control message within its diameter range. Any neighbour node that is already within its diameter range should be able to receive that message, otherwise, it will detect a hole at that point.

Phase III (coverage restoration): after detection of the hole by the $s_i$ node, that node begins to estimate a new sensory range and gains its distance to the farthest node that comes out of the cover (node c).

$$\mathbf{D(s_i, c)} = \sqrt{(\mathbf{x_i} - \mathbf{x})^2 + (\mathbf{y_i} - \mathbf{y})^2} \tag{1}$$

Considering the residual energy of each node, they have a chance to increase their sensory range. In other words, if a node has more residual energy, the node has a higher chance to cover the area. Thus, after increasing the range of that node, it sends a control message and alerts the neighbour nodes to the new range. If the hole is fully covered, the rest of the nodes cancel the process of increasing their range. Our proposed method avoids the occurrence of the energy hole by utilizing mobile sensor nodes and adding an energy balance algorithm to [4]. The advantage of our proposed method compared to the previous methods is that by finding the most congested areas and providing suitable coverage on them. Moreover, the energy balance in the network is determined by the target motion. This means where more power is needed, more energy and more nodes need to be present. For this purpose, after estimating the frequency of moving targets in different areas of the network, the areas with more traffic will be covered with more nodes.

The organization of this paper is as follows. In Section 2 research background and routing methods are described. In Section 3 the proposed algorithm is explained. In Section 4 the evaluation of the proposed method and its comparison with other method are carried out and eventually in section 5, the conclusion is expressed.

## 2 Related work

In the literature of hole coverage problem, Nguyen [1] introduced new algorithm based on graph theory In both sensor modes that are static and mobiles for coverage hole healing problem to improve remaining energy for mobile sensors. In [2], the whole of network is partitioned into grids. After calculating the coverage rate of all grids, grids are that their coverage rate are lower than others. Then by using mobile sensors, coverage hole is determined. In this article[3], the authors proposed an algorithm that include two step. First, they detect the coverage hole. Then, the coverage hole is restored. Hadikhani et al. [4] proposed a distributed algorithm for dynamic hole coverage to reduce the energy consumption. Initially, they divide the network into the unit grid squares by using a two-dimensional array. Next, they find the boundary of hole approximately with the help of stuck sensors. Afterwards, by using sensors which called pivot that are located in the each grids, the hole coverage is updated. In [5], a connectivity based k-coverage hole detection algorithm is presented. A reduction algorithm is proposed to reduce the size of the network topology. After simplification, initially, the boundary of 1-coverage holes is Identified. Then, by detecting independent sensors in the covered areas and turning them to sleep, the coverage degree reduces. Ma et al. [6] uses geometric approach to detect the coverage holes and their goal is to surge energy consumption. The advantages of their method is that region with various shape is kept under observation. In [7], a primitive topological approach is presented which can detect the range of internal and boundary holes of wireless sensor networks. This is a simple distribution method that identifies the location of nodes near the boundary of the hole boundary. Regarding Dividing and conquer Algorithms in [8-10] which are an efficient way of dividing the problem that solve the hole coverage problem by splitting it. Hence, by covering the holes of the sensor network into smaller components and solving the problem in each part of it, the computational complexity decreases. Yari et al. [11] introduced an Energy-efficient topology to enhance the wireless sensor network. They used two meta-heuristic algorithms to increase the energy remaining in the sensors. Moreover, a low-cost spanning tree was utilised to create an appropriate connectivity control among nodes in the network in order to increase the network lifetime.

## 3  Problem Description and Contribution

Full coverage of network quickly destroys the network because of the cost of node spreading and physical limitations. As a result, the energy balance among the nodes, increasing the lifetime of the network and the appropriate coverage on the target area are the important issues that we are going to find solutions. In [1-3], they used mobile sensor nodes to cover points in the network that was out of the nodes' coverage for various reasons. In these papers, the mobile sensor nodes play a key role, so that whenever the energy hole is observed, the mobile nodes move toward it and cover the location of the hole. But it is clear that these methods are not applicable to wireless sensor networks with fixed nodes. In [4], a method for solving the black hole problem in wireless sensor networks was introduced. They used fixed nodes and by increasing their sensing range the holes around the nodes are covered. However, one of the problems in this system is that it requires a large number of sensors to cover the environment and support the sensors and the energy balance is not fully visible. Another problem in [4] is that not only a large number of sensors are not guaranteed to reduce power consumption but also more control messages must be sent and all static nodes are active.

The contribution of this paper is to a new approach is proposed to prevent and eliminate holes in wireless sensor networks and reduce energy consumption by combining the method presented in [4] and the use of mobile nodes and clustering of the network. An example of network clustering is shown in Figure 1. The purpose of the proposed method is to prevent the holes occurrence and detect the holes with the lowest energy consumption.

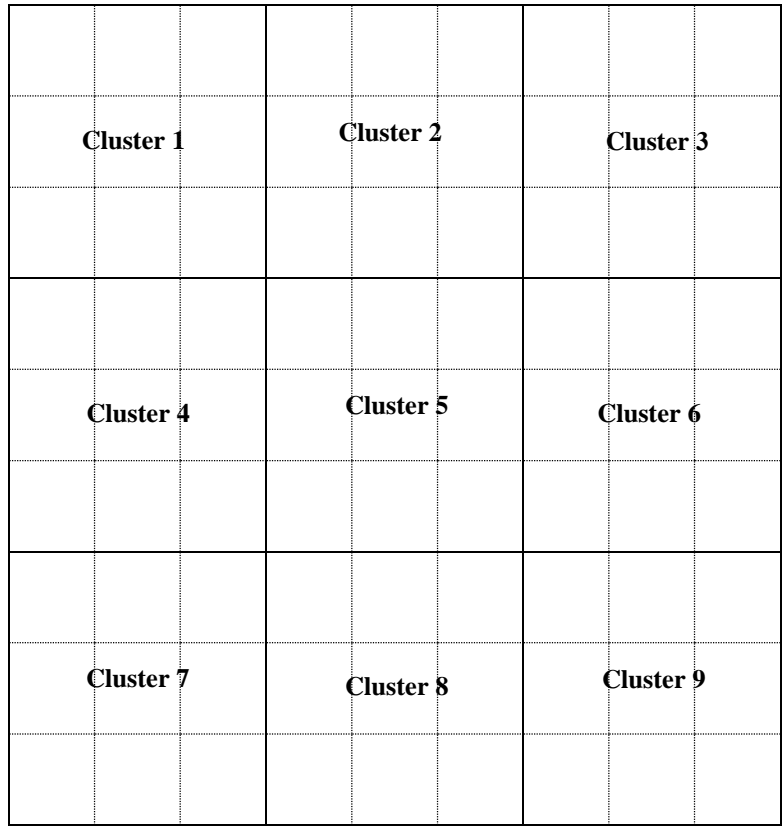

Figure 1: Clustering the network environment

# 4 Important terms

Some of the terms used in the design of the proposed method are discussed below:
- Target Area: If A is a target area, this region is divided based on a grid pattern with N × N squares. Each square is called a cell and for each cell, we consider Coordinates which are expressed as C (x, y). Figure 2 shows the cellularization of the environment:

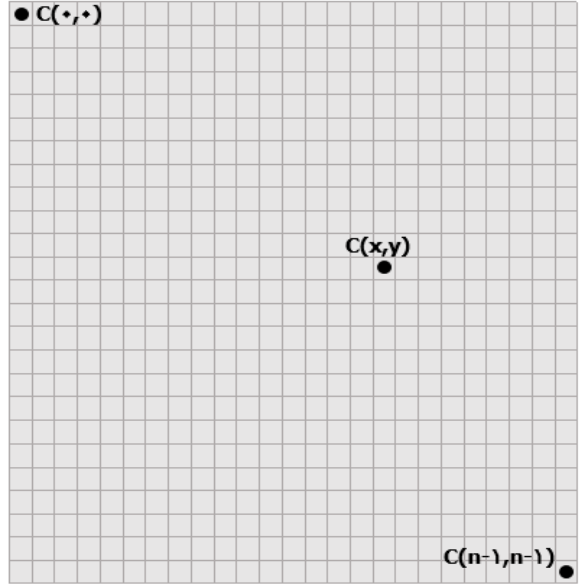

Figure 2: Division the target area into cells

- Full coverage: Suppose $S_1, S_2, \ldots, S_n$ are a set of fixed sensors in the network. The full coverage of the target region is recognized by the sensors using the following formula, where $S_{ij}$ is the region covered by the node j in the sub-regions $A_i$:

$$p = \frac{(A \cap (\cup_{i=1}^N A_i \cap (\cup_{j=1}^N S_{ij}))}{A} \tag{2}$$

Where N is the number of sensors in the network and $A_i \cap \left(\cup_{j=1}^N S_{ij}\right)$ is the set of cells in the $A_i$ region. If p = 1, it means that the whole area is covered and each cell in the target area is covered with at least one sensor. Figure 3 shows the full coverage of a sub-region in the network.

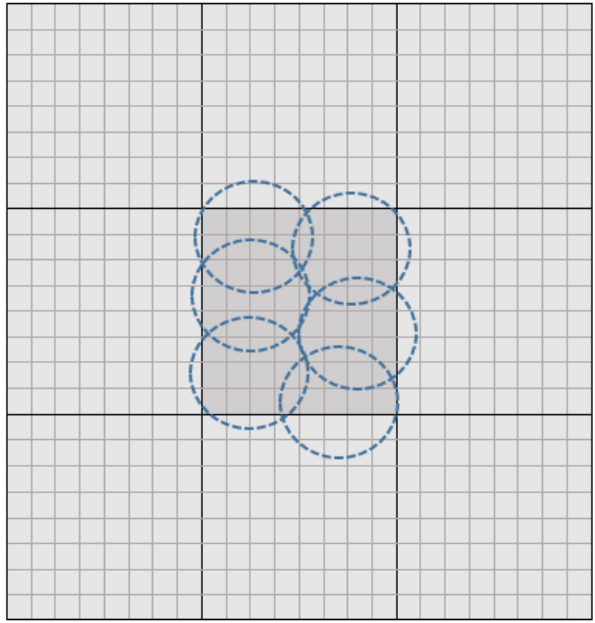

Figure 3: Full coverage of a sub-regions in the network. The small squares show the cells and the large squares show the sub-regions of network.

- Covered cell: A cell with geographical coordinates C (x, y) is covered by a $S_i$ sensor located at location C $(x_i, y_i)$, if:

$$C(x, y) \in A(S_i) \ if \ \sqrt{(x_i - x)^2 + (y_i - y)^2} \leq R_l(i) \qquad (3)$$

In this equation, $R_l(i)$ is the sensory range of the $S_i$ node. Figure 4 shows the covered cells.

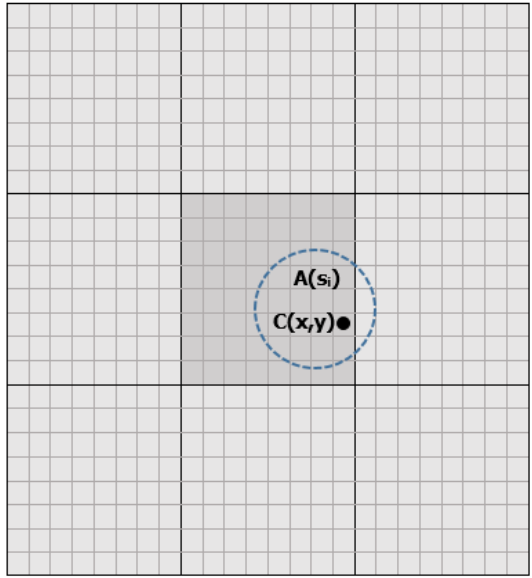

Figure 4: Covered cell

# 5 Assumptions

The network model in this study has the following assumptions:
- The nodes are uniformly positioned in a two-dimensional environment.
- Some nodes are fixed and some are moving.
- The energy of all nodes is the same and each node is aware of its location.
- On startup, the sensory range of the nodes is equal and is indicated by $R_l$.
- Nodes are able to increase the sensory range to $R_s$. Obviously $R_l<R_s$.
- There are a number of mobile backup nodes in the network that move to places where they are needed.

# 6 Proposed method

In this approach, first, holes are identified and covered locally in the clusters by collecting sensors' information and the heads of each cluster. The network is also periodically monitored by a sink node to process cluster information and detect existing holes. Moreover, the whole the network is divided into constant sub-regions. In each sub-regions, we select one of the nodes as the cluster head using the ER-HEED method which was proposed in [12]. These nodes send a message to all the nodes around them. This message contains the header ID and the header location information. Nodes that are not cluster head receive and record all information from the headers within their transmission range. The proposed method has four phases:
- Hole prevention phase
- Update phase
- Hole detection phase
- Hole Coverage phase

## 6.1 Hole Prevention Phase

In this phase, clusters that have recorded events or mobile Target have passed through them are monitored by the Cluster heads and the amount of energy and residual energy consumed by each segment of clusters are calculated. This information will be used in addition to statistical information about the occurrence of events in each cluster sent from the update phase to this phase to perform hole prevention operations. The primary energy of all nodes is considered equal and can be derived from the following energy consumption:
The power consumption of each sensor node to transmit the l-bit of information to a distance $d$ follows the following pattern:

$$E_{Tx}(l,d) = \begin{cases} l.E_{elec} + l.\varepsilon_{fs}.d^2 & if\ d < d_0 \\ l.E_{elec} + l.\varepsilon_{mp}.d^4 & if\ d \geq d_0 \end{cases} \qquad (4)$$

Where $E_{Tx}$ (l, d) is the energy used to transmit l-bit information to a distance $d$, where $d_0$ is the threshold of distance between two radio stations. The amount of energy consumed to receive the l-bit of data in the receiver also follows the equation 5.

$$E_{Rx}(l) = l \cdot E_{elec} \tag{5}$$

The residual energy of the sensors is obtained by subtracting the primary energy and the energy consumed (the sum of energy consumed for sending information and the energy consumed for receiving information). Each sensor transmits the residual energy to its cluster heads, and a threshold is empirically calculated for residual energy in the cluster. Because the sensors are location aware, they can also detect residual energy for different ranges of clusters.

The energy threshold can be calculated as the ratio of residual energy in a region's sensors to the total residual energy of the cluster. Here, we take the value of this ratio by 0/1. Naturally, this ratio is influenced by the time and energy factor in the cluster. If this ratio is less than 0/1, it means that there is a probability of a hole in that area of the cluster, and the characteristics of the area as a crisis zone will be sent to the detection phase to provide a mobile node for it.

## 6.2 Update Phase

In this phase, information on the energy of the network nodes and the statistics of the mobile nodes in the network and cells covered of each cluster are updated. First, the sensors inside a cluster find the cells in their range. The set of these cells is represented by $Q_l(i)$. If the Si node is at position C $(x_i, y_i)$, a line of $2 \times R_l(i)$ is drawn horizontally and vertically from location C $(x_i, y_i)$. Divide $R_l(i)$ into part H so that:

$$H = \left[\frac{r_s(i)}{l_c}\right] \tag{6}$$

The cells with the following coordinates place in the sensory range of the Si node:

$$\begin{aligned}
Q_{l1} &= \{(x,y)|(x_i + k_1 * l_c, y_i - k_2 * l_c)\}, k_1 = 0,1,2,\ldots,H-1 \\
Q_{l2} &= \{(x,y)|(x_i - k_1 * l_c, y_i - k_2 * l_c)\}, k_1 = 0,1,2,\ldots,H-1 \\
Q_{l3} &= \{(x,y)|(x_i - k_1 * l_c, y_i + k_2 * l_c)\}, k_1 = 0,1,2,\ldots,H-1 \\
Q_{l4} &= \{(x,y)|(x_i + k_1 * l_c, y_i + k_2 * l_c)\}, k_1 = 0,1,2,\ldots,H-1
\end{aligned}$$

For any value of $k_1$, the value of $k_2$ is from 0 to H-1. And also the value of $Q_l(i)$ is calculated as follows:

$$Q_l(i)=\left(\cup_{k=1}^{4}\left(Q_l(i)-\varphi(i)\right)\right) \quad (7)$$

Here is $\varphi_1 \subset Q_l(i)$, that is, a set of cells that are inside $Q_1(i)$ and are outside the range $R_l(i)$. Figure 5 shows the 4-fold regions around each node as well as cells outside the node range.

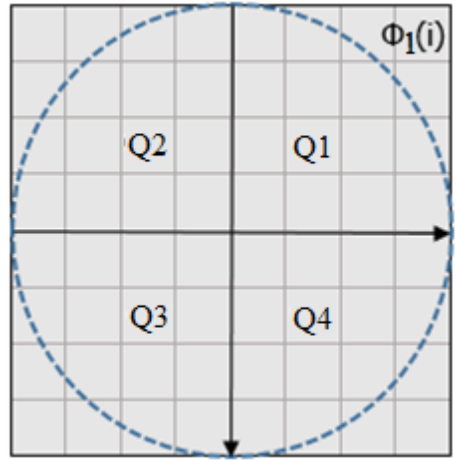

Figure 5:An example of how to find the set $Q_l$

The $S_i$ node can also find a set of cells that are within the $R_s$ coverage range, and the set of these cells lies in $Q^*(i)$. Similarly, the maximum area that can be located within the node can be identified and divided according to eequation 8.

$$H^* = \left\lceil \frac{R_s}{l_c} \right\rceil \quad (8)$$

The cells with the following coordinates are in the $R_s$ range of the $S_i$ node:

$$Q_1^* = \{(x,y)|(x_i + k_1 * l_c, y_i - k_2 * l_c)\}, k_1 = 0,1,2,\ldots,(H-1)^*$$
$$Q_2^* = \{(x,y)|(x_i - k_1 * l_c, y_i - k_2 * l_c)\}, k_1 = 0,1,2,\ldots,(H-1)^*$$
$$Q_3^* = \{(x,y)|(x_i - k_1 * l_c, y_i + k_2 * l_c)\}, k_1 = 0,1,2,\ldots,(H-1)^*$$
$$Q_4^* = \{(x,y)|(x_i + k_1 * l_c, y_i + k_2 * l_c)\}, k_1 = 0,1,2,\ldots,(H-1)^*$$

For any value of $k_1$, the value of $k_2$ is from 0 to (H-1). And the value of $Q_l(i)^*$ is calculated as follows:

$$Q^*(i)=(\cup_{k=1}^{4}(Q^*(i)-\varphi^*(i))) \quad (9)$$

The sensors then update the $Q_{l-s}$ set as follows:

$$Q_{l-s}(i) = Q^*(i) - Q_l(i) \tag{10}$$

Where $Q_{l-s}$ represents a set of cells outside the range $R_l(i)$ and inside the range $R_s$. The update phase flowchart is shown in Figure 6.

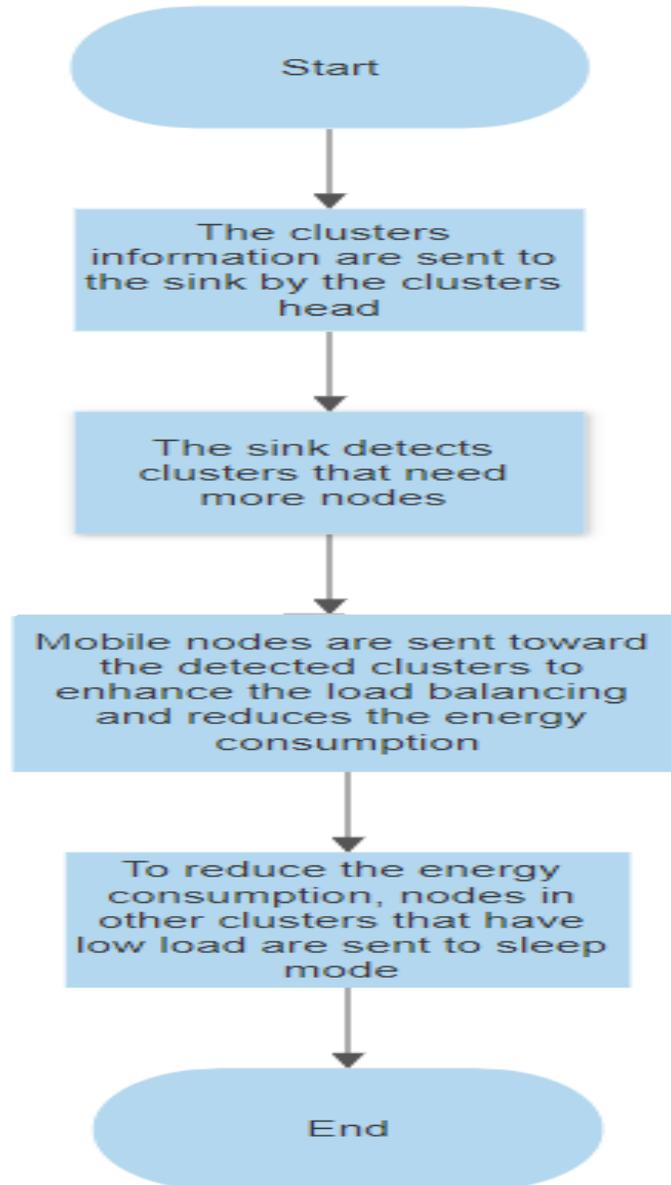

Figure 6: Global update flowchart

Whenever a global update is made, load balancing are calculated based on the energy consumption of cluster heads. In this approach, one counter in each cluster is responsible for recording the information. This information are collected by cluster heads and send toward the sink. These information are crucial for preventing energy hole. Given the proportion of occurrences in each cluster and the amount of energy consumed and residual energy, the sink can easily determine which clusters or regions need more nodes. Using information, the sink can send mobile nodes to high-load areas to enhance the load balancing and reduces the energy consumption. It is also worth noting that the counter of cluster heads are zeroed after submitting this information to the sink and resume their counting.

### 6.3 Hole Detection Phase

In this phase, each sensor node first transmits a control message to all its neighbouring nodes in the cluster with the range $R_c$ which is $R_c = 2R_s$. The message includes ID of the sensor node, location information and $Q_l(i)$. $N^*(i)$ is a set of all neighboring nodes in the range of $R_c$. If the $S_i$ node receives a control message from its neighbours ($N^*(i)$), the $S_i$ node calculates a new set $Q\hat{}(i)$ as follows:

$$Q\hat{}(i) = Q_{l-s}(i) - (Q_{l-s}(i) \cap (U_{s_k \in N^*(i)} Q_l(s_k))) \tag{11}$$

If $Q\hat{}(i) = \emptyset$, it means that the cells belonging to the set $Q'(i)$ are covered by neighbouring nodes of $S_i$. But if $Q\hat{}(i) \neq \emptyset$ means that there is a hole near $S_i$. Figure 7 shows the occurrence of the hole and Figure 8 shows the flowchart of the hole detection phase.

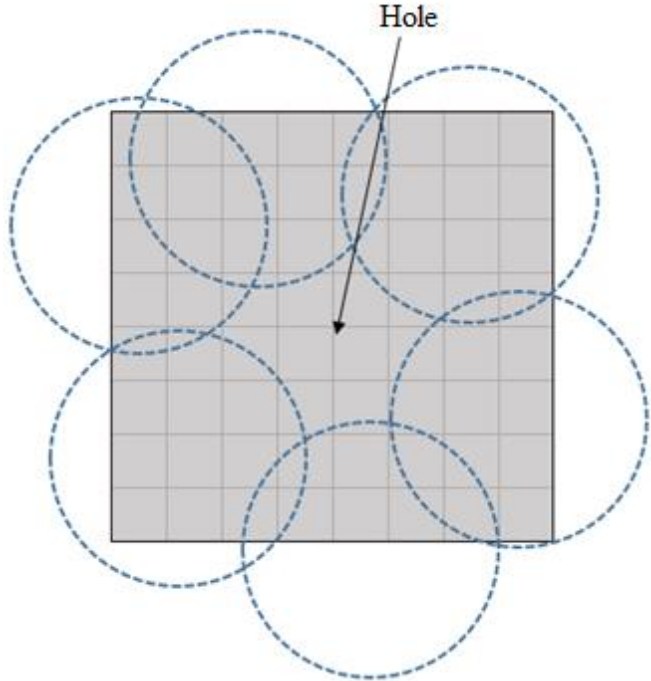

Figure 7: Hole Occurrence

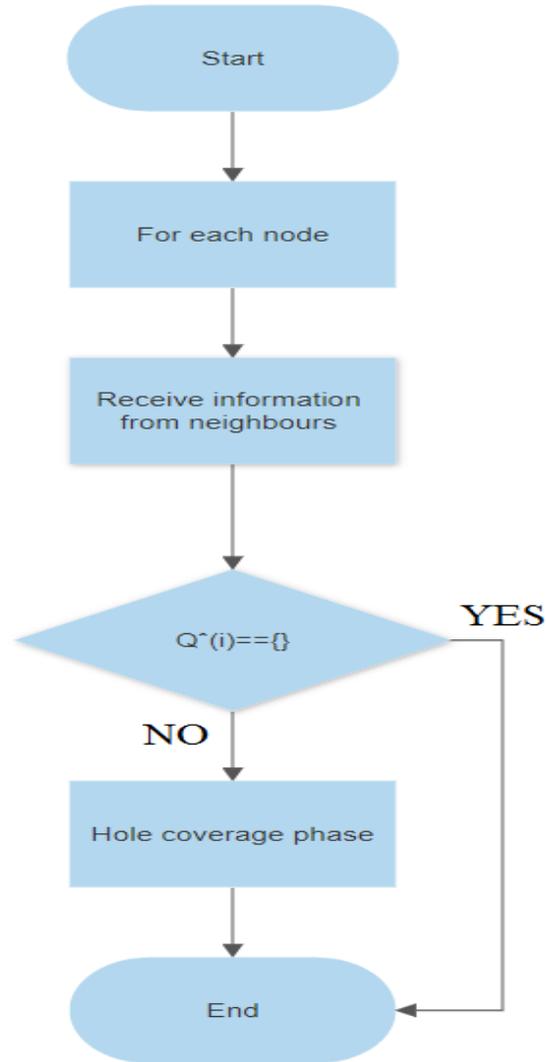

Figure 8: Flowchart Hole detection phase

## 6.4 Hole Coverage Phase

To cover the hole, initially, each nodes around the hole consider a timer for themselves according to their $Q\hat{}$ value. The higher the $Q\hat{}$, the less self-timer they have. Whenever the timer runs out, they recheck the hole and if they cannot covered the hole, they increase their sensory range and try to cover the hole. It should be noted that the timing of this timer is very short and also end after the hole is covered. Under these conditions, the small holes are easily covered. But if the sensory range was not able to cover the hole, they would use mobile nodes around them. If there is no mobile nodes around them, they will be assisted by their cluster head. The flowchart of this operation is illustrated in Figure 9.The existence of a timer causes the node that is closest to the hole to cover it. As a result, coverage is often accomplished by increasing the sensory range of a node. It also prevents unnecessary sending of control messages.

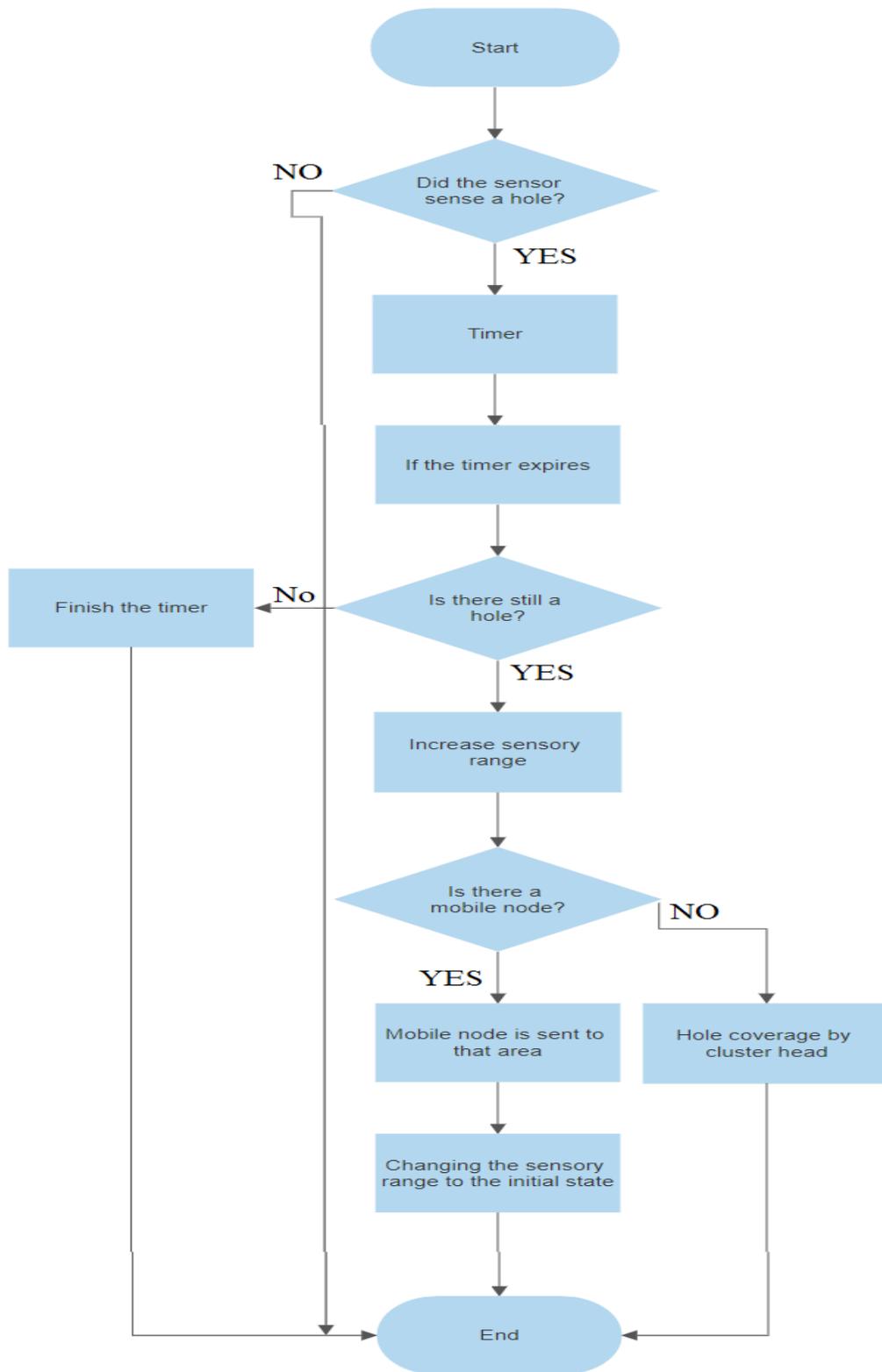

Figure 9: Flowchart cover hole by node

The clusters try to resolve the problem locally by increasing the sensory range of all of their sensors that cover the hole. But if they are unable to solve the problem, they will request help from adjacent clusters. When Help requests are sent to the neighbour cluster, only the mobile nodes will be sent to assist those areas and there will be no increase in sensory range. The Request cluster nodes immediately decrease their sensory range and return to their normal state after resolving the hole Coverage by mobile nodes. These steps are revealed by Figure 10.

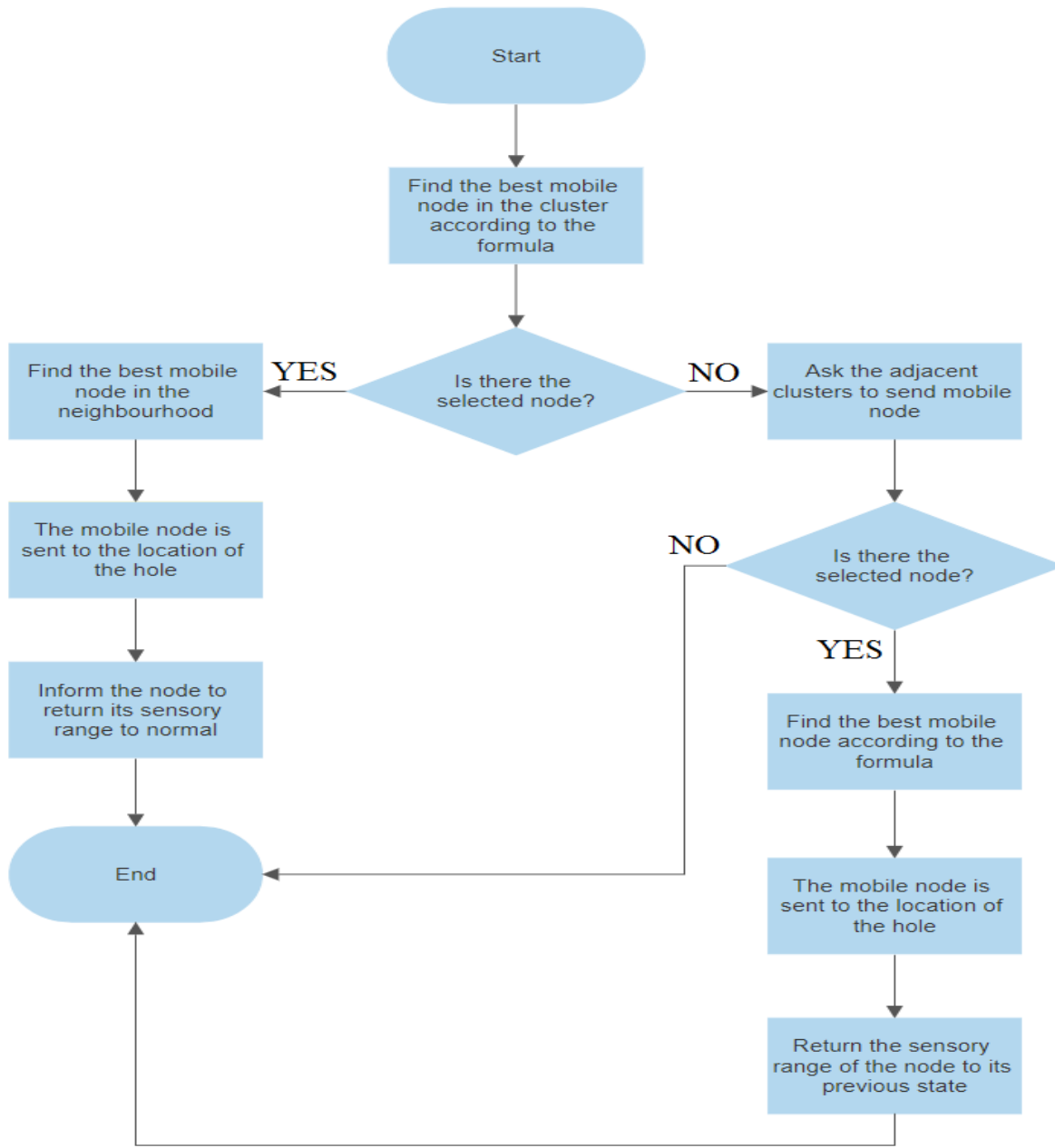

Figure 10: Flowchart of Hole Coverage by cluster head and neighbor clusters

In this phase, the distance of the mobile nodes to the hole is considered as main factor to choose the best mobile node. Mobile nodes send a messages to the cluster head. Therefore, after identifying the hole, the cluster head selects from among the mobile nodes which have less distant to the hole. Since all mobile nodes are asleep from the beginning, they have the same initial energy, so their residual energy is not a criterion for choosing a better node. If we show mobile nodes with $SM_j$, the selection criterion for each mobile node is calculated by the relation 12:

$$SM_j = distance \tag{12}$$

Then the best mobile node to get into the hole location is obtained from the following formula:

$$SM_{best} = max(distance) \tag{13}$$

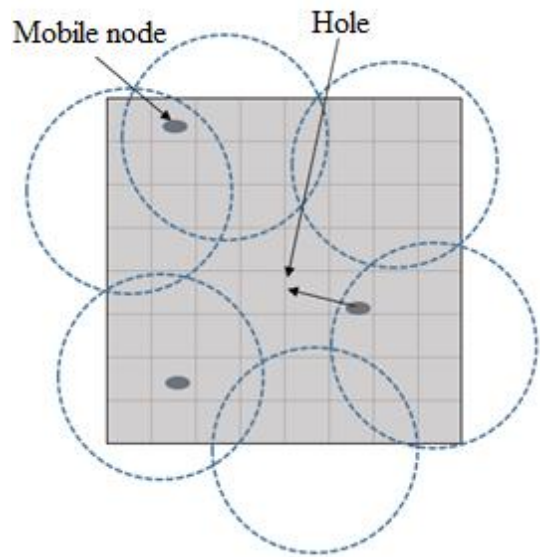

Figure 11: Moving the mobile nodes towards the hole

Figure 11 shows how the moving node moves to the hole. After selecting the best mobile node using the equations 12 and 13, the node moves to the hole location and sets the flag option to one. (That is, this mobile node is used in the network and no longer used in the selection of the mobile node for other area).

Figure 12 also shows the state of the network after the node is moved to the hole location.

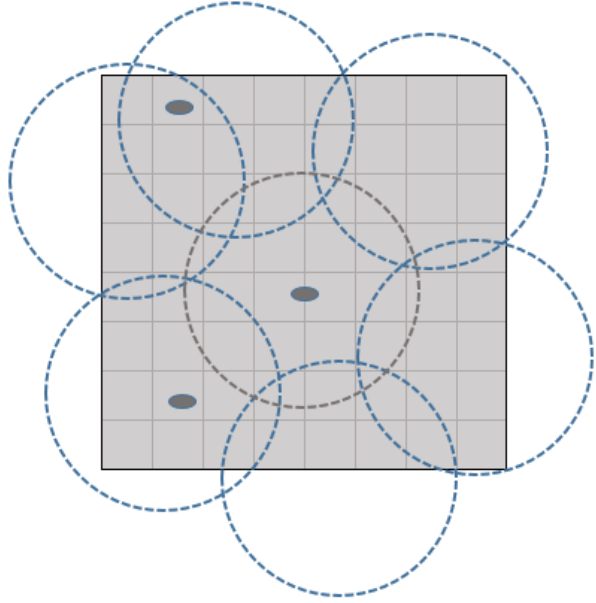

Figure 12: Hole coverage by mobile node

# 7 Simulation and Results

In this section, we use the NS-2 network simulator to simulate the proposed method and the basic article. The simulation parameter of the first experiment is given in the table below. Also, the number of runs to obtain the average data is 25 times.

Table1: simulation parameters

| Initial energy of each node | 4J |
|---|---|
| Network Dimensions | 2000m*2000m |
| Node communication range | 200m |
| Sensory radius of the node | 20m |
| Number of node | 200,300,400,500 |
| Position of nodes | randomly |
| $E_{amp}$ | 0.01J |
| $E_{trans}$ | 0.02 |
| $E_{receive}$ | 0.01J |
| Data packet size | 512 bytes |
| Target Motion Model | Random way point |
| Percentage of nodes breakdown | 10 |
| Times of simulation | 2000s |

## 7.1 Energy Model

The energy consumed by the nodes includes the energy used to store and the energy used to retrieve data. The energy required to transmit according to equation 14 and the energy required to receive the data by the node on the basis of equation 15. The energy model presented is as follows:

$$E_{send} = E_{trans} * s + E_{amp} * d^2 \tag{14}$$

$$E_{receive} = E_{recv} * s \tag{15}$$

$E_{trans}$ is the amount of energy consumed for transmitting a data bit, $s$ is message size (each packet), $d$ is message transmission distance. $E_{amp}$ is The energy consumed in the signal amplifier, $E_{receive}$ is the energy required to receive one bit of data.

## 7.2 Experiment and analysis

Experiments and analysis of the results are based on energy consumption criteria, load balance, hole coverage lifetime and recovery time.
- Energy Consumption Criterion: Energy is a major issue in the design of a sensor wireless network system. Because each sensor node has limitation on the amount of energy and they cannot replace this energy.
- Load Balance Criteria: Load balancing is a model for distributing load across a network or components of a network such as links, nodes, and users to achieve maximum resource utilization and maximum throughput in the shortest possible time. Also, load balancing plays an effective role in preventing overhead and providing quality assurance while multiple nodes are considered for the same task.
- Hole coverage lifetime Criterion: The concept of Hole coverage means covering the holes created during the lifetime of the network It is calculated based on how long the holes in the network are covered. If the hole coverage time is longer, the quality of service in the network will improve.

- Recovery Time: The length of time the network detects a hole in the normal state and then the hole is covered, is called the recovery time (return to the normal state of the network).

## 7.3 Number of nodes

In this section, the influence of changes in the number of nodes is checked in the simulation. For this purpose, the number of nodes is set to 200, 300, 400 and 500 and the results are compared with the basic algorithm based on 4 criteria of energy consumption, load balance, coverage life and recovery time (in third scenario only).

- The effect of increasing number of nodes on energy consumption

    The average energy consumption increases slightly as the number of nodes increases. The reason for this is the control messages sent between the nodes, which increase with the number of nodes as the number of these messages increases. Another reason for this upward trend is the consumption of nodes in sleep mode. In our proposed method, gathering information during the updating and hole prevention phases by control messages makes the energy consumption increases due to the increasing number of nodes. In the case of hole, the higher the number of nodes, the greater the number of nodes used for the cover operation and the more control messages are transmitted. but the reason for the superiority of the proposed method is based on choosing the best sensory range for nodes as well as the phase of preventing the hole occurring makes reduce the need for the rescue operation, and consequently, the fair distribution of energy improves the energy consumption and stability of the network even if balanced power consumption and network stability in the event of a malfunction.

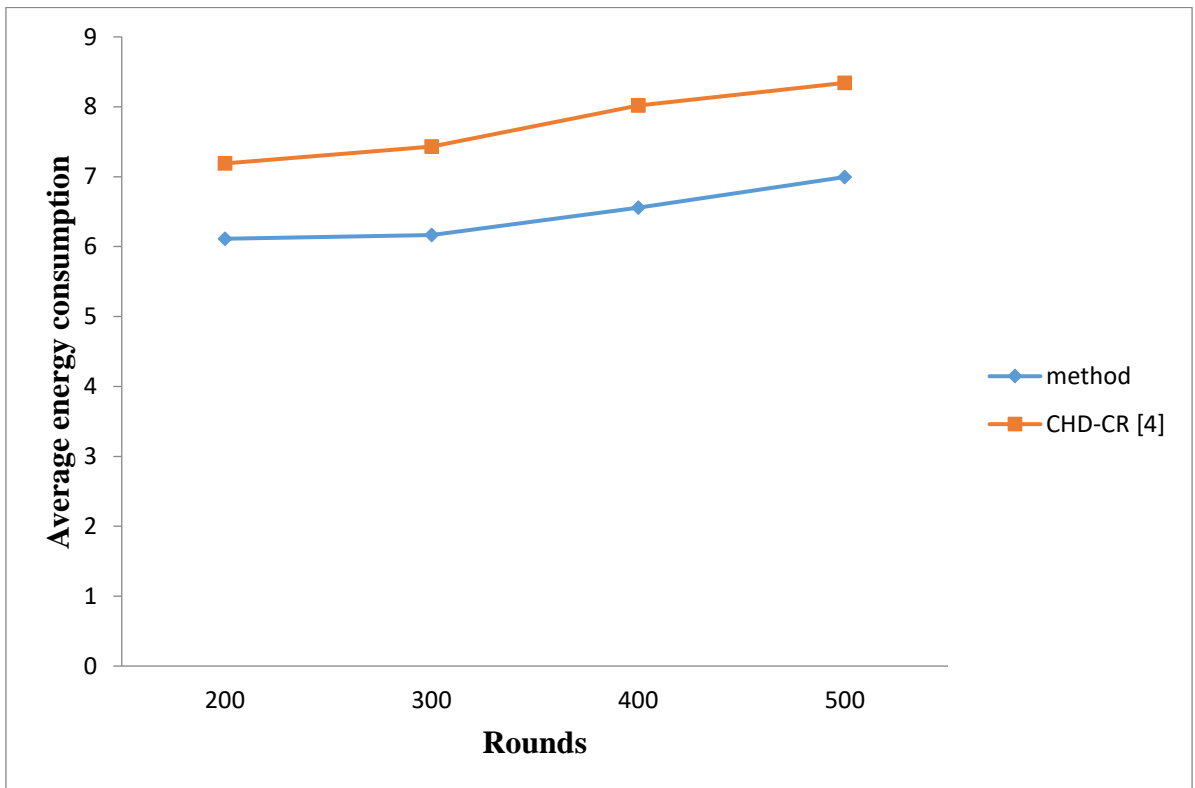

Figure 13: Energy consumption versus number of nodes

- The effect of increasing number of nodes on lifetime of the hole coverage

    The Figure 14 indicates an increase in energy consumption when the number of nodes increases and also because of the support nodes as well as the moving nodes, the lifetime of the hole coverage in the proposed method is better than the base method.

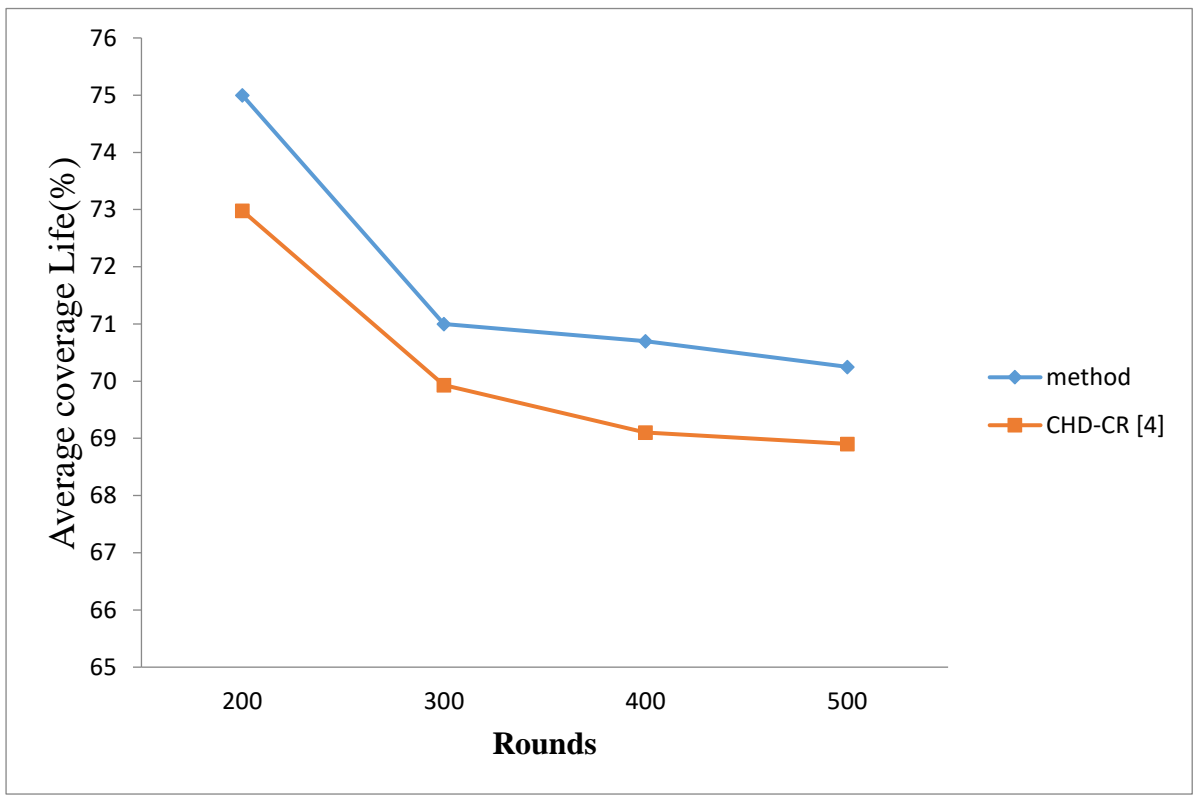

Figure 14: The effect of increasing number of nodes on lifetime of the hole coverage

- Influence of node number on average load balance

Load balancing means a fair distribution of computational, sensory and sending data among the network nodes in such a way that nodes completely cover the interest area and consume nearly the same energy consumption. Due to this reason, Load balancing increases network lifetime and reduces sudden energy depletion. Figure 15 reveals the influence of node number on average load balance. We can see that as the node increases, the Load balancing decreases. On the contrary, it does not seem to be a bad thing. In the both method, only a few nodes are used to keep the network well covered and others to fall asleep. Although the energy of both of them has decreased, in the proposed method has better Load balancing due to detection of crowded spots and data transmission among the network nodes.

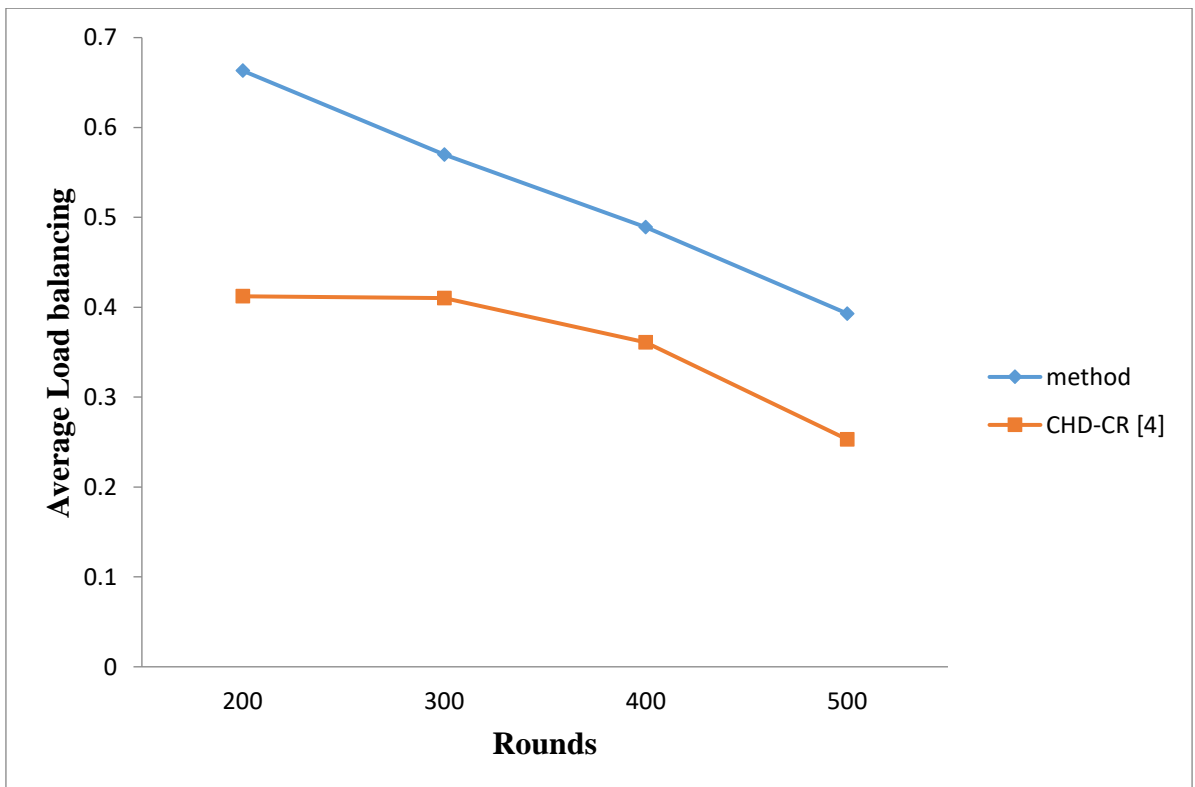

Figure 15: Influence of node number on average load balance

## 7.4 Node failure

In this section, we investigate the effect of node failure percentage on the performance of the proposed method. The number of nodes is set to 100 and the results are compared with the basic algorithm based on energy consumption, load balance, coverage life.

- Impact of node failure on energy consumption

    In this experiment, we destroyed 25, 50, and 75% of the nodes, respectively, to measure the energy consumption and network stability. The results show that with increasing failure rate in both algorithms the average energy consumption increases because the higher the failure rate, the more routing it will require. Moreover, energy consumption increases because the increased sensory range of fixed nodes and the motion of mobile nodes to cover holes.

Table2: simulation parameters

| Parameter | Value |
|---|---|
| Node initial energy | 5J |
| Network Dimensions | 1000m*1000m |
| Node communication range | 200m |
| Sensory radius of the node | 20m |
| Number of node | 100 |
| Position of nodes | randomly |
| $E_{amp}$ | 0.1J |
| $E_{trans}$ | 0.1J |
| $E_{receive}$ | 0.1J |
| Data packet size | 512byte |
| Target Motion Model | Random way point |
| Percentage of nodes breakdown | 25,50,75 |
| Times of simulation | 1000s |

The reason why the proposed method is superior to the basic article method is using mobile nodes. However, in [4], as the sensory range increases, the nodes lose their energy faster. Therefore, lower energy consumption is used in the proposed method compared to [4].

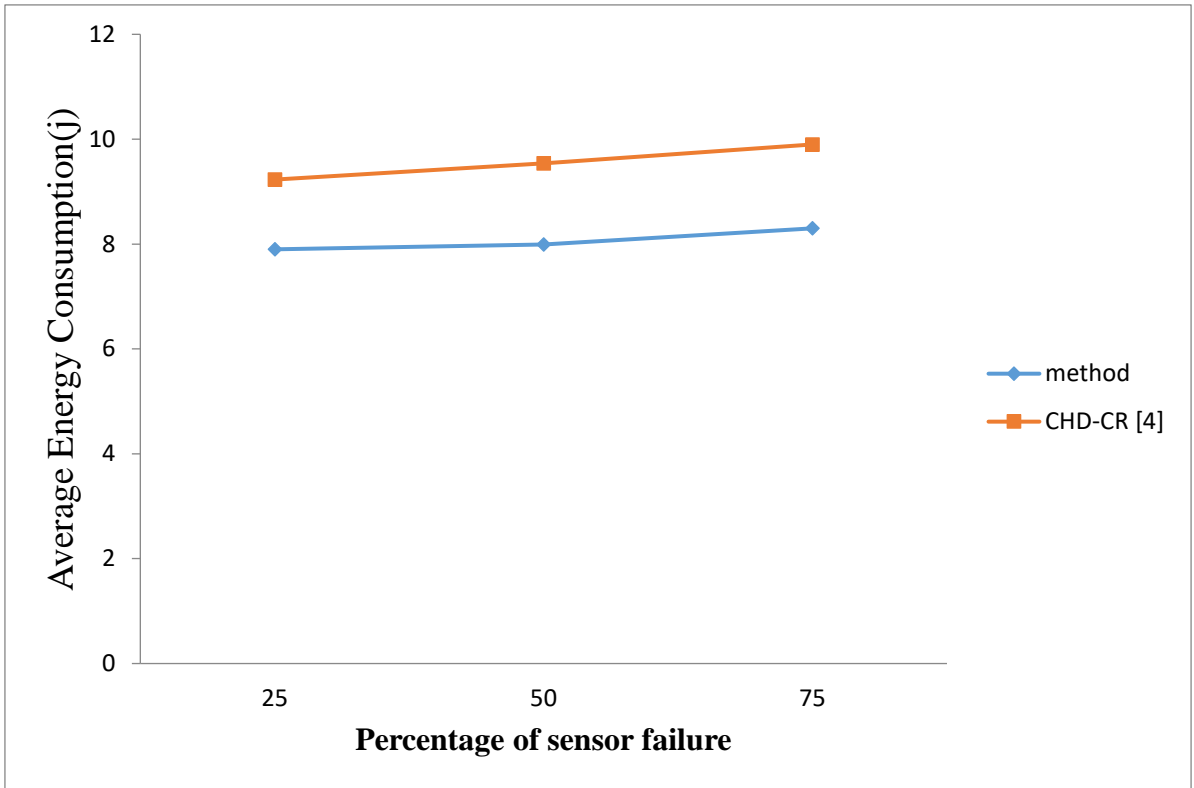

Figure 16: Impact of node failure on energy consumption

- The effect of node failure on lifetime of the hole coverage

    In this case, as the failure rate increases, the hole coverage will become more difficult. But, due to the hole coverage phase, Updating environment and event logs in all clusters in the proposed method, the lifetime of the hole coverage has improved. As the failure rate increases, support nodes are turned on or mobile nodes are sent to critical points and fill holes.

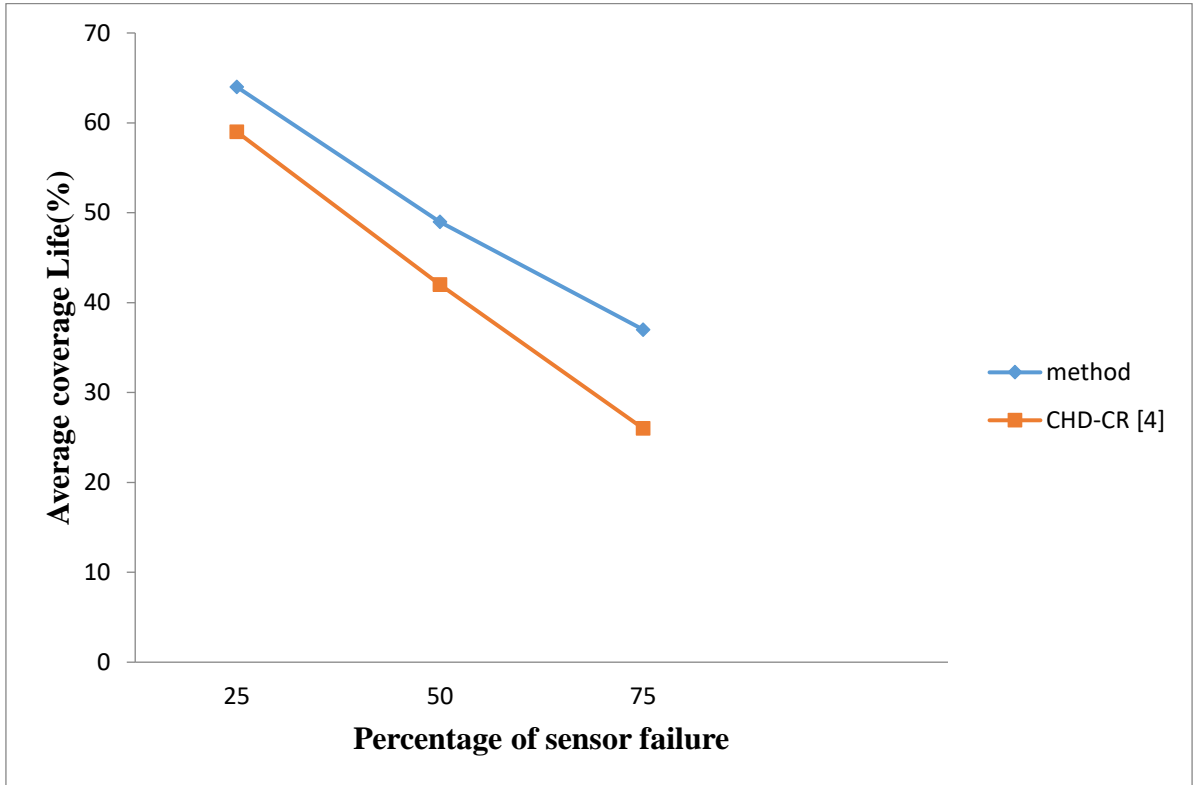

Figure 17: The effect of node failure on lifetime of the hole coverage

- Influence of node number on average load balance

    As shown in Figure 18, the proposed method has better load balancing than other method due to use of mobile nodes. In this way, after the nodes of an area are destroyed, the mobile nodes are sent to that area to keep the network balanced.

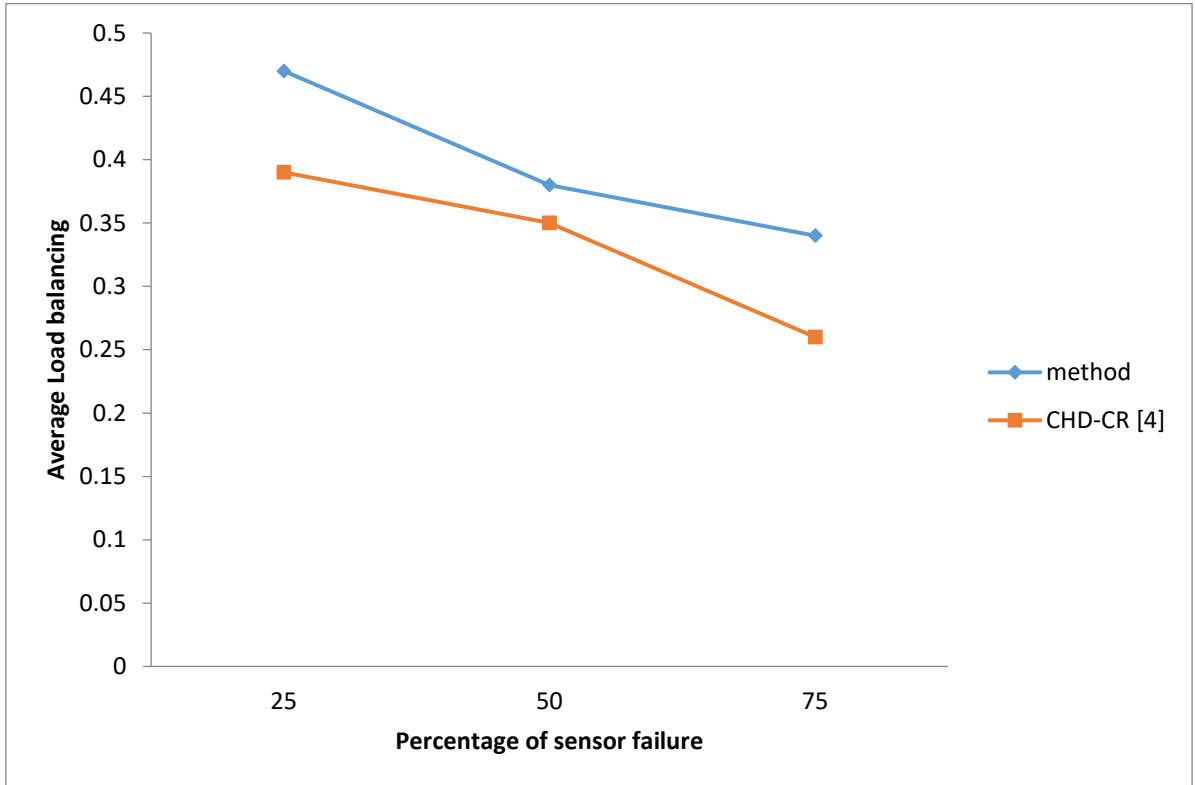

Figure 18: Influence of node number on average load balance

# 8   Conclusion

The results of the simulations show that the proposed method, in addition to reducing energy consumption, has been able to increase the balance and lifetime of the hole coverage. The solution applied in the long run when very large holes occur, it can act as an independent repair algorithm (AuR) [13] and apply the necessary coverage to the environment. Statistics obtained from the multiplicity of events in different locations make the network more stable over the long run and reduce the likelihood of holes. Using the variable Sensory radius length, we chose the most appropriate interval which each node consumes less energy and reduces the energy consumption of the entire network. Using the equitable distribution of energy and resources among different regions, we came closer to the goal of energy balance. In this way, active nodes that produce more data are supported by more nodes in order to backup them when node crash or Energy discharge node happened. Moreover, long-term global updates will return the network to a unified state.

As the future works, we would like to explore hybrid algorithm like [14] to improve the performance of the generative model. We would also like to extend our algorithm by using connectivity control [15] to improve load balancing among the network nodes.

# 9  REFERENCES

1. Nguyen DT, Nguyen NP, Thai MT, Helal A, "An optimal algorithm for coverage hole healing in hybrid sensor networks", 2011.
2. Li F, Xiong S,Wang L, "Recovering coverage holes by using mobile sensors in wireless sensor networks", 2011.
3. Abo-Zahhad M, Ahmed SM, Sabor N, Sasaki S, " Rearrangement of mobile wireless sensor nodes for coverage maximization based on immune node deployment algorithm", 2015.
4. Tarachand Amgoth, Prasanta K. Jana, "Coverage hole detection and restoration algorithm for wireless sensor networks", 2015.
5. Yasir Faheem, Saadi Boudjit, Ken Chen. Dynamic Sink Location Update Scope Control Mechanism for Mobile Sink Wireless Sensor Networks. 2011.
6. Wang, J., et al., A mobile assisted coverage hole patching scheme based on particle swarm optimization for WSNs. Cluster Computing, 2019. 22(1): p. 1787-1795.
7. Amgoth, T. and P.K. Jana, Coverage hole detection and restoration algorithm for wireless sensor networks. Peer-to-Peer Networking and Applications, 2017. 10(1): p. 66-78.
8. Hadikhani, P., et al., An energy-aware and load balanced distributed geographic routing algorithm for wireless sensor networks with dynamic hole. Wireless Networks: p. 1-13.
9. Yan, F., et al., Connectivity Based k-Coverage Hole Detection in Wireless Sensor Networks. Mobile Networks and Applications, 2019: p. 1-11.
10. Ma, H.-C., P.K. Sahoo, and Y.-W. Chen, Computational geometry based distributed coverage hole detection protocol for the wireless sensor networks. Journal of network and computer applications, 2011. 34(5): p. 1743-1756.
11. Yari, M., P. Hadikhani, and Z. Asgharzadeh, Energy-efficient topology to enhance the wireless sensor network lifetime using connectivity control. arXiv preprint arXiv:2005.03370, 2020.
12. Ullah, Zaib & Mostarda, Leonardo & Gagliardi, Roberto & Cacciagrano, Diletta & Corradini, Flavio. (2016). A Comparison of HEED Based Clustering Algorithms -- Introducing ER-HEED. 339-345. 10.1109/AINA.2016.87.
13. Yatish K. Joshi, Mohamed Younis, " Autonomous recovery from multi-node failure in Wireless Sensor Network" 2012 IEEE Global Communications Conference (GLOBECOM), DOI 10.1109/GLOCOM.2012.6503187.
14. Hadikhani, P. and P. Hadikhani, An adaptive hybrid algorithm for social networks to choose groups with independent members. Evolutionary Intelligence, 2020: p. 1-9.
15. Yari, M., Hadikhani, P., & Asgharzadeh, Z. (2020). Energy-Efficient Topology to Enhance the Wireless Sensor Network Lifetime Using Connectivity Control. Journal of Telecommunications and the Digital Economy, 8(3), 68-84.